\documentclass[fleqn,11pt]{wlscirep}

\usepackage{bm}

\usepackage{color}

\newcommand{\dif}{\mathrm{d}}

\usepackage{graphicx}
\usepackage{multirow}
\usepackage{amsmath}
\usepackage{amsfonts}
\usepackage[version=3]{mhchem}
\usepackage{graphicx}
\usepackage{breqn}
\usepackage{soul}

\title{Molecular Dynamics of Lithium Ion Transport in a
Model Solid Electrolyte Interphase}

\author[1]{Ajay Muralidharan}
\author[2]{Mangesh I. Chaudhari}
\author[1]{Lawrence R. Pratt}
\author[2,*]{Susan B. Rempe}
\affil[1]{Tulane University, Department of Chemical and Biomolecular Engineering, New Orleans, 70118, USA}
\affil[2]{Sandia National Laboratories, Center for Biological and Engineering Sciences, Albuqueruque, 87185, USA}

\affil[*]{slrempe@sandia.gov}

\begin{abstract}
Li$^+$ transport within a solid electrolyte interphase (SEI) in lithium
ion batteries has challenged molecular dynamics (MD) studies due to
limited compositional control of that layer. In recent
years, experiments and {\em ab initio} simulations have identified
dilithium ethylene dicarbonate (Li$_2$EDC) as the dominant component of
SEI layers. Here, we adopt a parameterized, non-polarizable MD force
field for Li$_2$EDC to study  transport characteristics  of
Li$^+$ in this model SEI layer at moderate temperatures.
The observed correlations are consistent with recent MD results using a
polarizable force field, suggesting that this non-polarizable model is
effective for our purposes of investigating Li$^+$ dynamics over long
time scales. Mean-squared displacements distinguish three distinct  Li$^+$
transport regimes in EDC --- ballistic, trapping, and diffusive. Compared to liquid ethylene carbonate (EC), the nanosecond trapping times
in EDC are significantly longer and naturally  decrease at
higher temperatures. New materials developed for fast-charging Li-ion batteries should have smaller trapping regions. The analyses implemented in this paper can be used for testing transport of Li$^+$ ion in novel battery materials.  Non-Gaussian features of van Hove 
\emph{self}-correlation functions for Li$^+$ in EDC, along with the mean-squared
displacements, are consistent in describing  EDC as a glassy material
compared with liquid EC. Vibrational modes of Li$^+$ ion, identified by MD, characterize the trapping and are further validated by electronic structure calculations. 
\end{abstract}

\begin{document}
\flushbottom
\maketitle

\thispagestyle{empty}
 
\section*{Introduction}
During charging and discharging cycles of lithium ion batteries, a solid
electrolyte interphase (SEI) layer forms  on the negative electrode due
to decomposition of solvents 
like ethylene carbonate (EC). The SEI
layer is a complex organic material and its composition is not
operationally set.\cite{NatureSEI2015} Nevertheless, dilithium ethylene
dicarbonate (Li$_2$EDC) has been identified experimentally as the
primary component of the outer part of the SEI
layer.\cite{identify_edc,Zorba:2012hc,Xu:2007ie,Leung:2016jq}

$Ab~initio$ molecular dynamics (AIMD) simulations, electronic structure
calculations\cite{BalbuenaP,LeungK,Borodin:2016gm} and reactive force
field simulations\cite{Kim:2011dn} on the decomposition of ethylene
carbonate (EC) on anode surfaces concur with those experimental results.
Experimental observations also show that the SEI layer protects the
electrode from further decomposition by blocking electron transport
while simultaneously allowing transport of Li$^+$ ions between the electrode and
electrolyte solution. A better understanding of the  transport mechanism
 of Li$^+$ in EDC may lead to modified SEI layers with improved lithium
ion battery performance. 
\begin{figure*}\centering
	\includegraphics[width=3.25in]{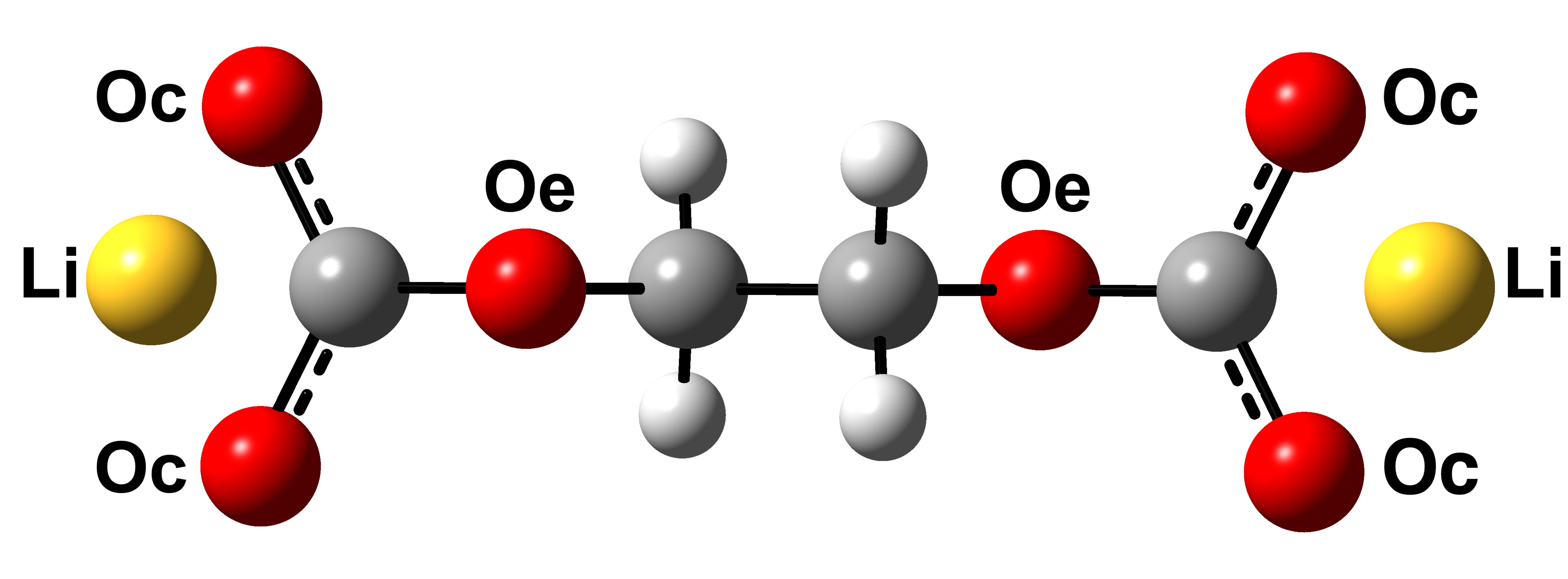}
	\includegraphics[width=2in]{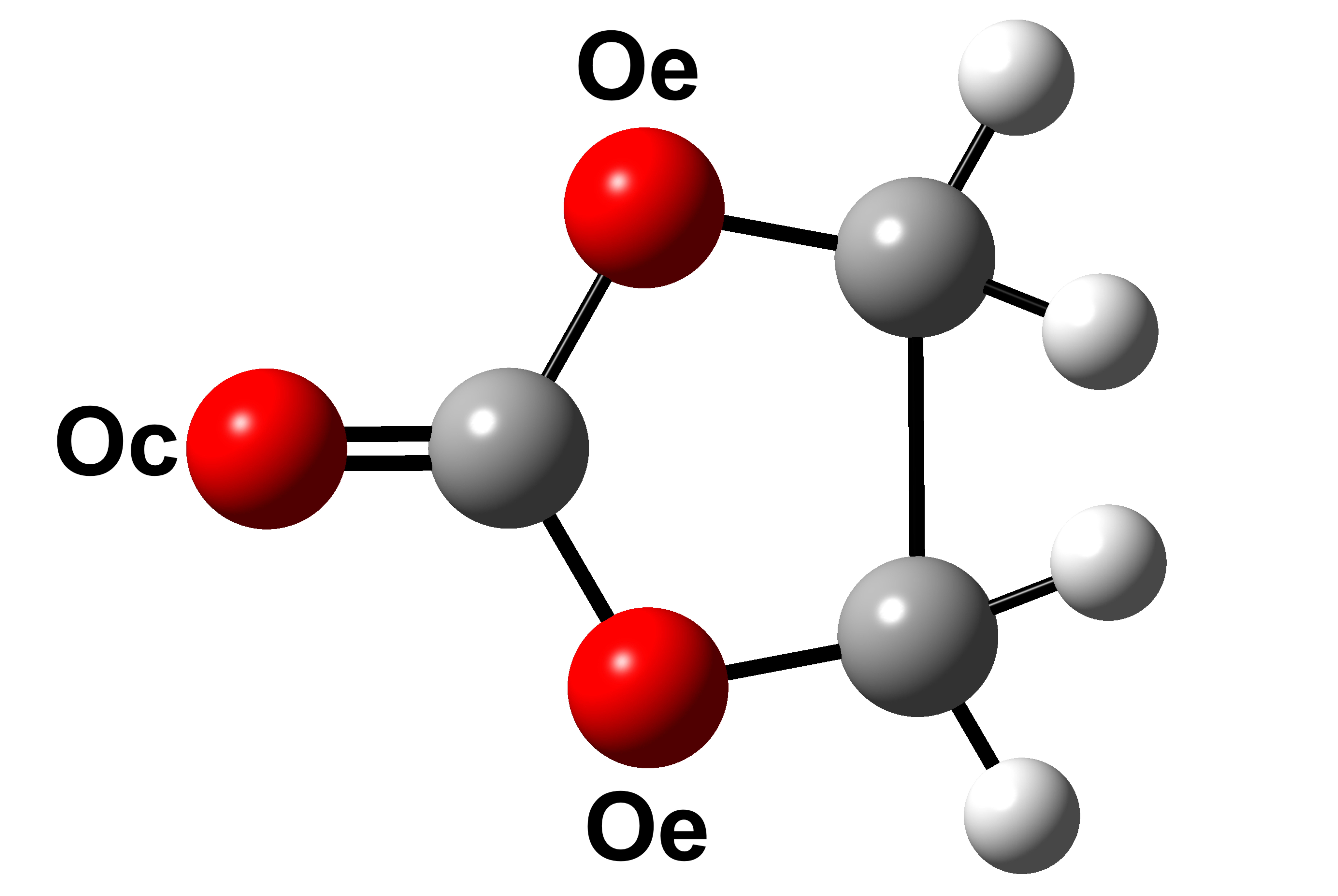}
	\caption{Chemical structure of Li$_2$EDC (left) and EC (right)
	molecules. Carbon atoms are colored gray, hydrogens white, carbonyl
	(Oc) or ether (Oe) oxygen red, and Li$^+$ ions
	yellow.}\label{li2edc}
\end{figure*}

Molecular dynamics (MD) studies on model SEI layers carried out over
long time-scales may shed new light into the mechanism of transport of
Li$^+$ ions within the SEI layer.  Borodin, \emph{et
al.,}\cite{Borodin:2013bq,Bedrov:2017ci,Jorn:2013kq} performed MD
calculations using a specialized polarizable force field to obtain 
transport properties of Li$^+$ ion in a model SEI layer composed of
ordered and amorphous Li$_2$EDC. Since polarizable force fields are not
readily available in standard molecular dynamics packages, we have
instead identified non-polarizable force field
parameters\cite{muralidharan2017molecular} for simulation of the
Li$_2$EDC model of the SEI layer. The microsecond time-scales studied
here, longer than earlier work,\cite{Borodin:2013bq} provide additional
insight into structural and transport properties of Li$^+$ ions in this
model SEI.

\begin{figure*}
\includegraphics[width=7in]{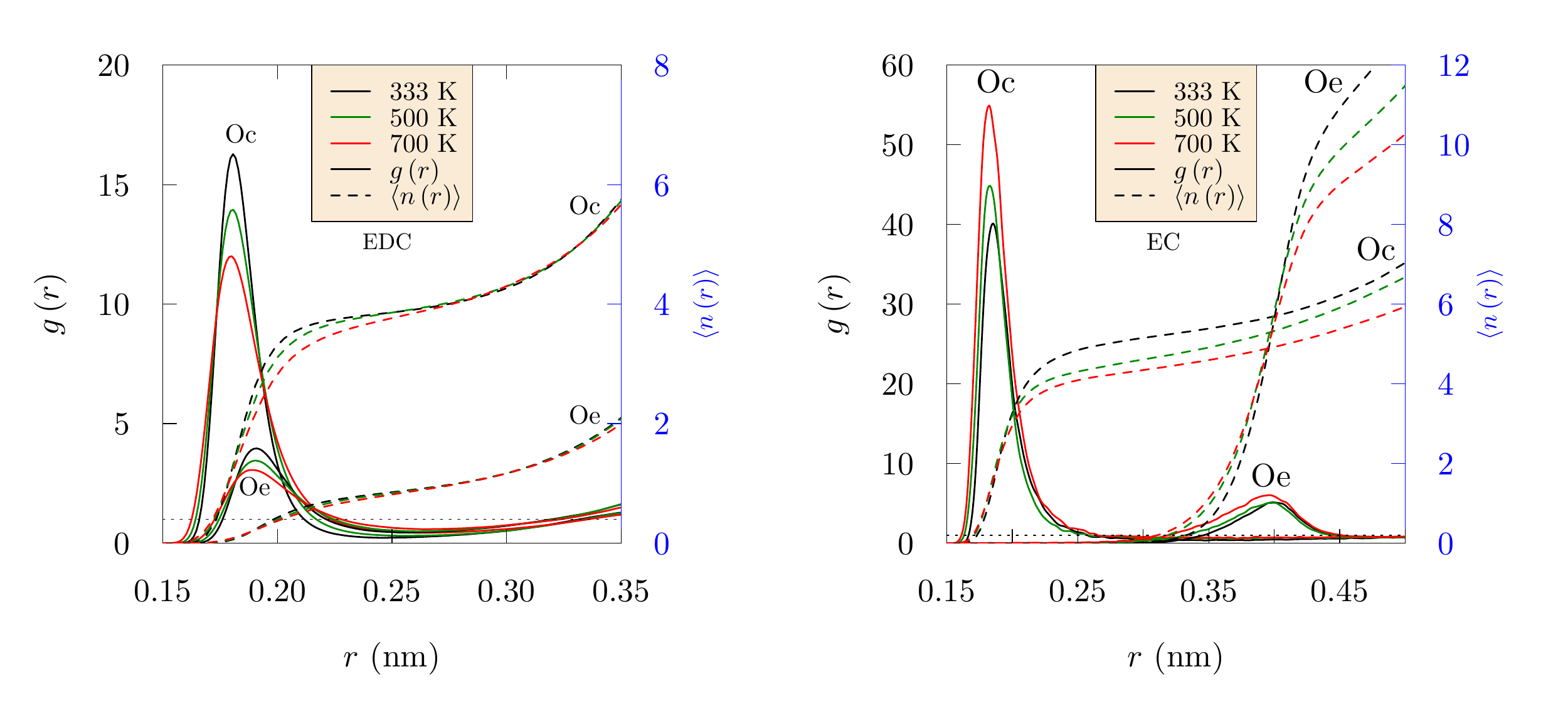}
\caption{Radial distribution functions, $g(r)$ and running coordination number,
$\langle{n(r)}\rangle = 4 \pi \rho  \int_0^r g(x) x^2 dx$,  for Oc-Li$^+$ and Oe-Li$^+$ at various
temperatures. For EDC (left), occupancy of the first solvation shell does not
depend on the temperature, even though the peak height diminishes with
increasing temperature. For liquid EC solvent (right), almost one additional Oc atom interacts with Li$^+$
at lower temperature.}
\label{fig:rdf_all} \end{figure*}

\section{Results and Discussion}
We summarize the structural and transport properties from MD studies of a model SEI layer with 256 Li$_2$EDC moieties (Figure~\ref{li2edc}). 
In addition, simulations of a dilute Li$^+$ ion in EC (single Li$^+$ ion solvated by 249 EC) provide a perspective for comparison. 
\subsection{Structural data}
The radial distribution functions (rdfs) and running coordination
numbers  (Figure~\ref{fig:rdf_all}), involving Li$^+$, carbonyl oxygen
(Oc) and ether oxygen (Oe) of EDC and EC are compared at several temperatures. For EDC, the rdfs become less structured with
increasing temperature, but the peak positions and overall coordination
numbers of the first peak change only slightly. This structural
robustness suggests an amorphous glassy matrix for the SEI. These
results compare well with recent polarizable force field
results,\cite{Borodin:2013bq,Borodin:2014gq}  supporting the
applicability of the present non-polarizable force field for these
structural characteristics. In EDC, the peak position for the Li-Oe
distribution is shifted to a slightly larger value than Li-Oc because
the charge on Oe is 40\% smaller than that of Oc. 

In the case of EC, the peak height increases with temperature and the
peak position of the Li-Oe is farther out due to strong interactions
between Li$^+$ and Oc. The structure around Li$^+$ changes significantly
with temperature, as highlighted by the running coordination number. In contrast to glassy EDC, almost one additional carbonyl oxygen (Oc) atom interacts with Li$^+$
at lower temperature for EC solvent.   

\subsection{Mean-squared displacements of Li$^+$}
\begin{figure*}
\includegraphics[width=7in]{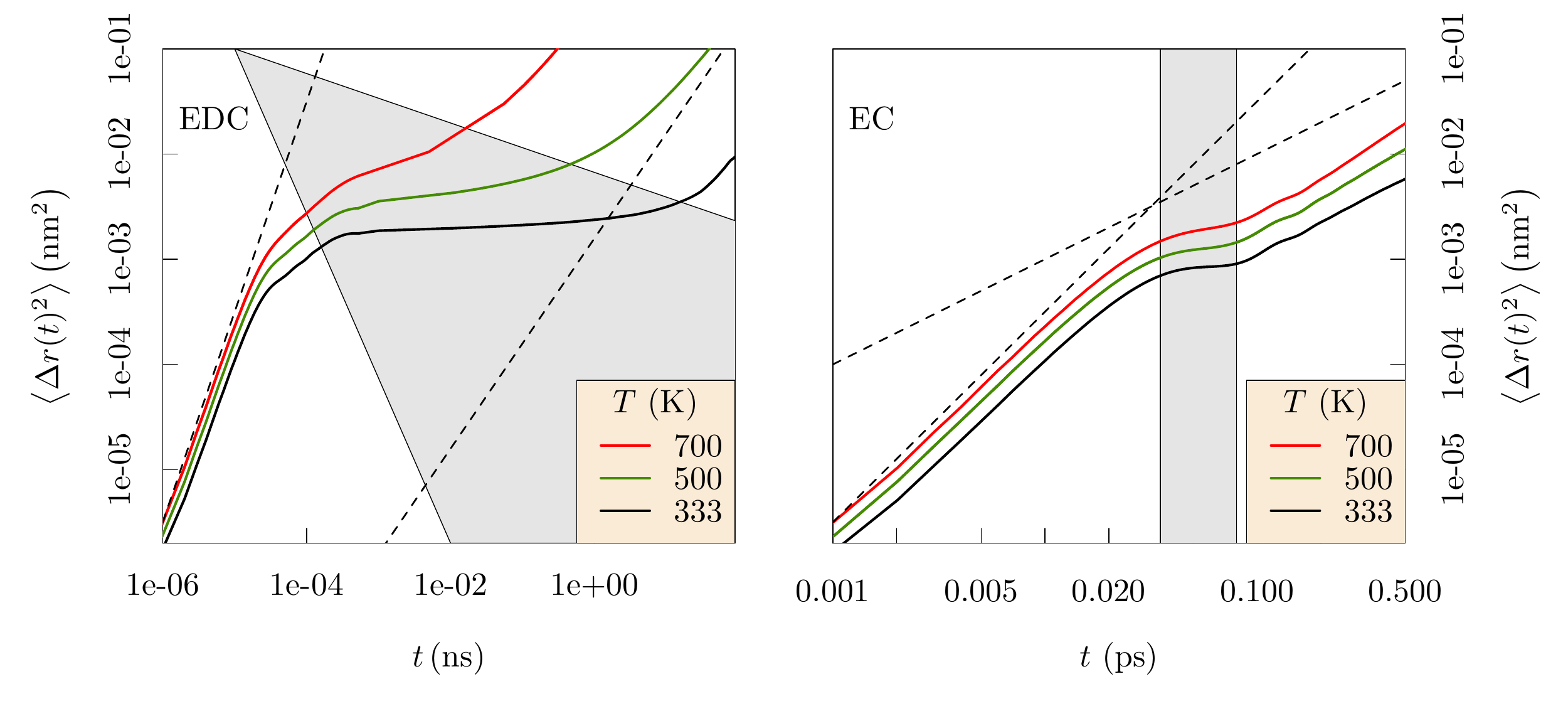}
\caption{Mean-squared displacements, $\langle\Delta r(t)^2\rangle$,
measured for Li$^+$ in EDC (left) and EC (right). The behavior in EDC at intermediate
timescales $0.001 < t < 1$ (ns) demonstrates trapping of the Li$^+$ ion.
Ballistic motion ($\langle\Delta r(t)^2\rangle \propto t^2$) is evident
at short timescales, while diffusive motion ($\langle\Delta
r(t)^2\rangle \propto t^1$) appears at long timescales in both EDC and EC
solvents. Dashed lines with slope 1 and 2 (log-scale) are provided as visual cues. At high $T$, the trapping regime (shaded region) diminishes and the EDC matrix behaves
more like liquid EC. Note that the time scales differ dramatically between
the two systems.} \label{fig:msdTCompositeTIkZ} \end{figure*}

The mean-squared displacements (MSD, Figure~\ref{fig:msdTCompositeTIkZ})
of Li$^+$ ion in EDC and liquid EC distinguish three distinct dynamical regimes:
ballistic at short times, trapping at intermediate times, and diffusive
at long times. Trapping of Li$^+$in EDC is qualitatively different than in 
 liquid EC, and the trapping times in EDC diminish as temperature increases and the 
glassy EDC matrix softens.
Diffusivities of  Li$^+$ are extracted from the slope of the diffusive regime of MSD and then extrapolated to
low temperatures using an Arrhenius fit (see Supplementary Information). The diffusivity of Li$^+$ in EDC at
333~K is 10$^{-12}$ cm$^{2}$/s, which is in agreement with Borodin \emph{et.~al.}\cite{Borodin:2013bq} The conductivity of Li$_2$EDC obtained from the Nernst-Einstein equation (Supplementary Information) is 4.5 x 10$^{-9}$ S/cm, which is also in agreement with experiment.\cite{Borodin:2013bq}

\begin{figure*}
\includegraphics[width=7in]{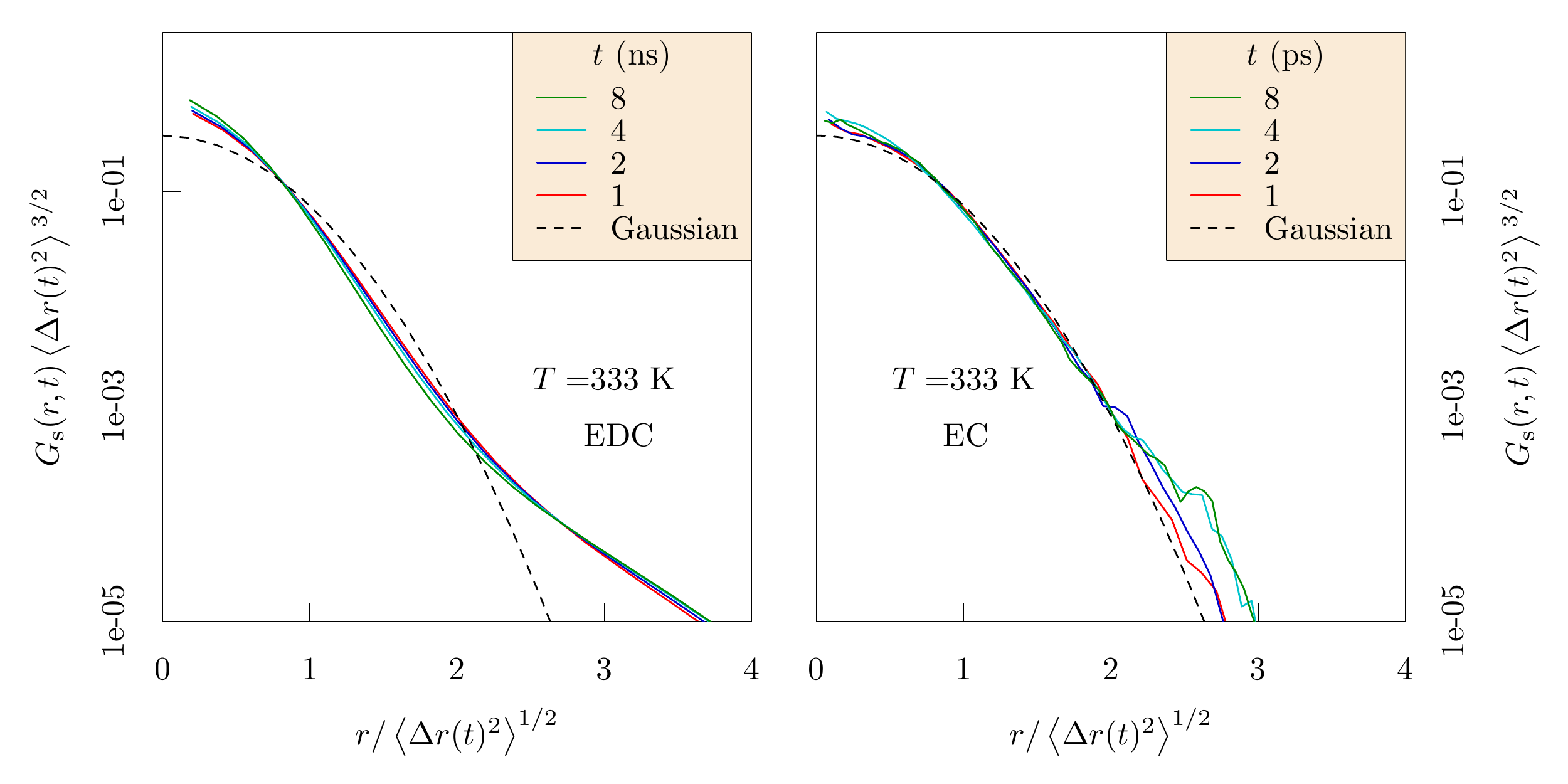}
\caption{Dimensionally-scaled $G_{\mathrm{s}}(r,t)$ for Li$^+$ ions as
it depends on displacement for increasing times in EDC (left) and EC
(right). Though the Gaussian model (dashed curve) is reliable in EC solvent, 
probability is depleted near the trap boundaries, $r> \langle\Delta
r(t)^2\rangle^{1/2}$, and replaced at shorter and longer distances for EDC. Note
that these correlations decay in a few ps for EC, but require ns for
EDC.}
\label{vH__333_RSTikZ}
\end{figure*} 

\subsection{Time correlation functions for Li$^+$ transport}
The van Hove time  correlation function
\begin{equation}
	G(r,t)={\frac{1}{N}}\left\langle\sum_{i=1}^N\sum_{j=1}^N\delta(\textbf{{r}}+\textbf{{r}}_j(0)-\textbf{{r}}_i(t))\right\rangle
\end{equation} 
for Li$^+$ ion describes the probability of finding, at time $t$, a
Li$^+$ ion displaced by $r$ from its initial position.
This
van Hove function can be split into \emph{self} and \emph{distinct} parts, $G(r,t)=G_{\mathrm{s}}(r,t)+
G_{\mathrm{d}}(r,t)$, with
the latter taking the form
\begin{equation}    
    G_{\mathrm{d}}(r,t)={\frac{1}{N}} \left\langle\sum_{i\neq{j}}^N(\textbf{{r}}+\textbf{{r}}_j(0)-\textbf{{r}}_i(t))\right\rangle~.
\end{equation}
At $t=0$, the van Hove function reduces to the static pair correlation function,
\begin{equation}
	G(r,0)=\delta({\textbf{r}}) + \rho g({r})~.
\end{equation}

The \emph{self} part of the van Hove function provides a jump
probability, and the natural initial approximation is the Gaussian model,
\begin{equation}
	G_{\mathrm{s}}(r,t)= \left\lbrack\frac{3}{2\pi \langle\Delta r(t)^2\rangle}\right\rbrack^{3/2} 
	\times \exp\left\lbrack - \left(\frac{3r^2}{2\langle\Delta{r(t)^2\rangle}}\right)\right\rbrack~.
\end{equation}
For fluids like EC, this Gaussian behavior should be reliable.  In
contrast, $G_{\mathrm{s}}(r,t)$ in the trapping regime of EDC indicates
(Figure~\ref{vH__333_RSTikZ}) depletion of probability near the trap
boundaries $r> \langle\Delta r(t)^2\rangle^{1/2}$, and replacement of
that probability at shorter and longer distances.  Deviation from the
Gaussian behavior can be characterized by the non-Gaussian
parameter\cite{Vorselaars:2007dg}
\begin{equation}\centering
	\alpha(t)= \frac{3\langle\Delta{r(t)^4}\rangle}{5\langle\Delta{r(t)^2\rangle^2}}-1 .
\end{equation}

\begin{figure}\centering
	\includegraphics[width=3.5in]{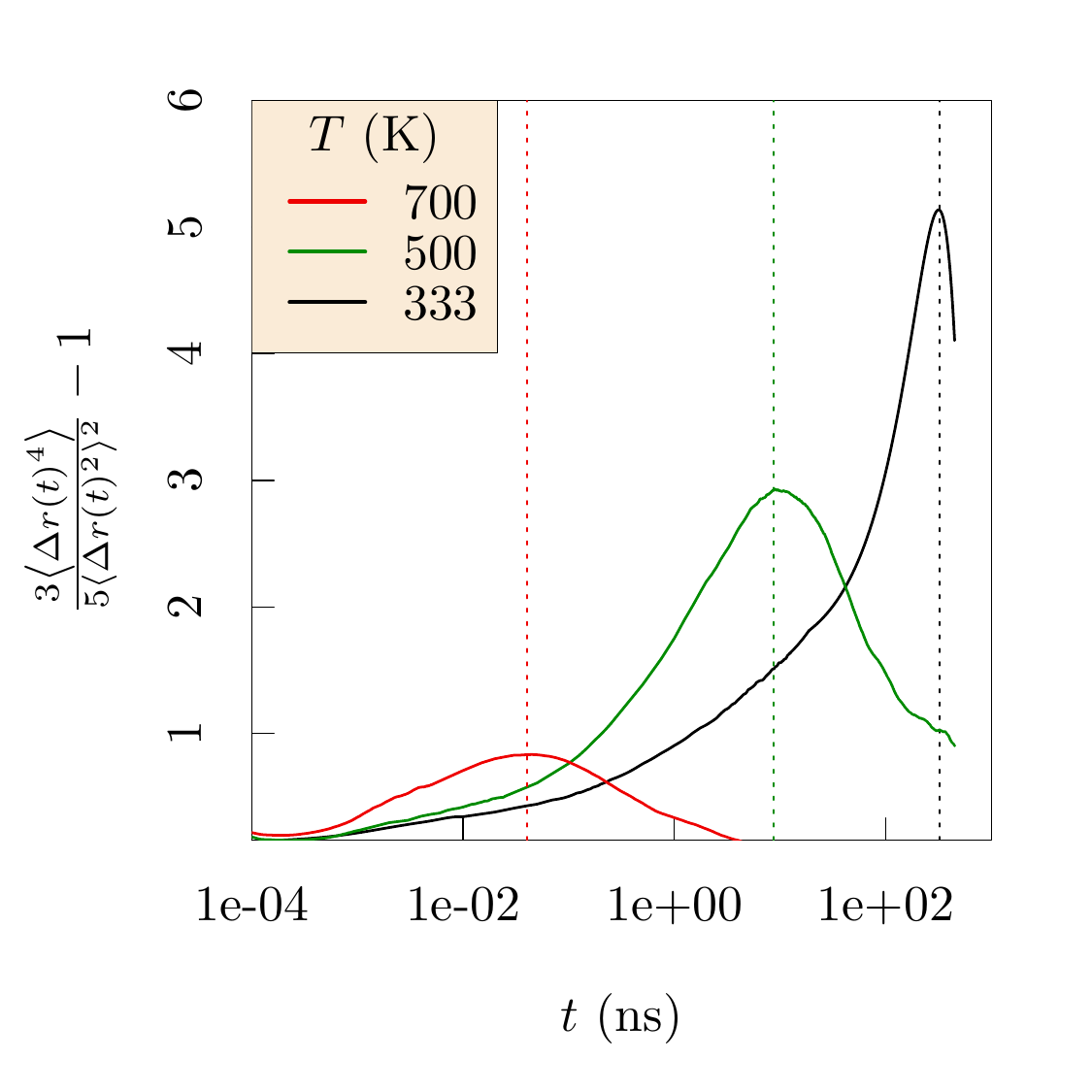}
	\caption{Non-Gaussian parameter calculated for Li$^+$ in EDC at several
	temperatures. Vertical lines are drawn at the maximum value of
	$\alpha(t)$. The non-zero value and inverse temperature dependence of
	$\alpha(t)$ attests to the glassy behavior of EDC, which becomes more
	fluid-like at higher temperature.}\label{fig:NGP}
\end{figure}

\begin{figure}\centering
	\includegraphics[width=3.5in]{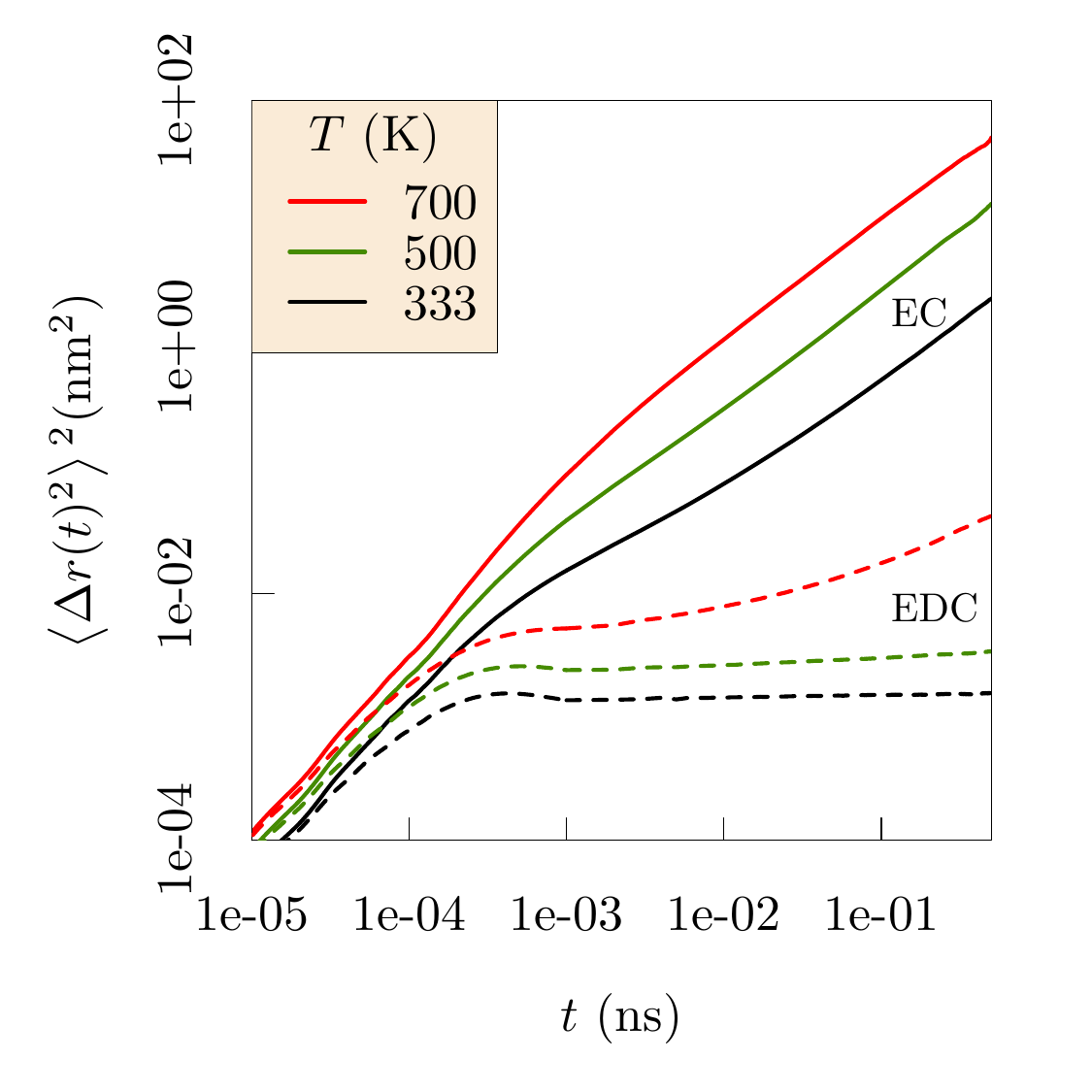}
	\caption{Comparison between mean-squared displacements 
	of
	carbonyl carbon of EDC and EC. The flat MSD at larger time-scales
	reflects the glassy nature of the EDC matrix.}\label{fig:MSD}
\end{figure}

The van Hove self-correlation function is accurately Gaussian for liquid
EC, hence $\alpha(t)=0$. In contrast, $\alpha(t)$ has non-zero values for glassy EDC
(Figure~\ref{fig:NGP}). For EDC, $\alpha(t)$ has a maximum  that decreases
with increasing temperature.  The mean-squared displacements of the
carbonyl carbons of EDC and EC molecules (Figure~\ref{fig:MSD}) further
verify the sluggish diffusion of EDC relative to EC. In the case of EDC, the
mean-squared displacement curves are relatively flat for all
temperatures, indicating little diffusion of the EDC matrix.   

\begin{figure}
\includegraphics[width=3.5in]{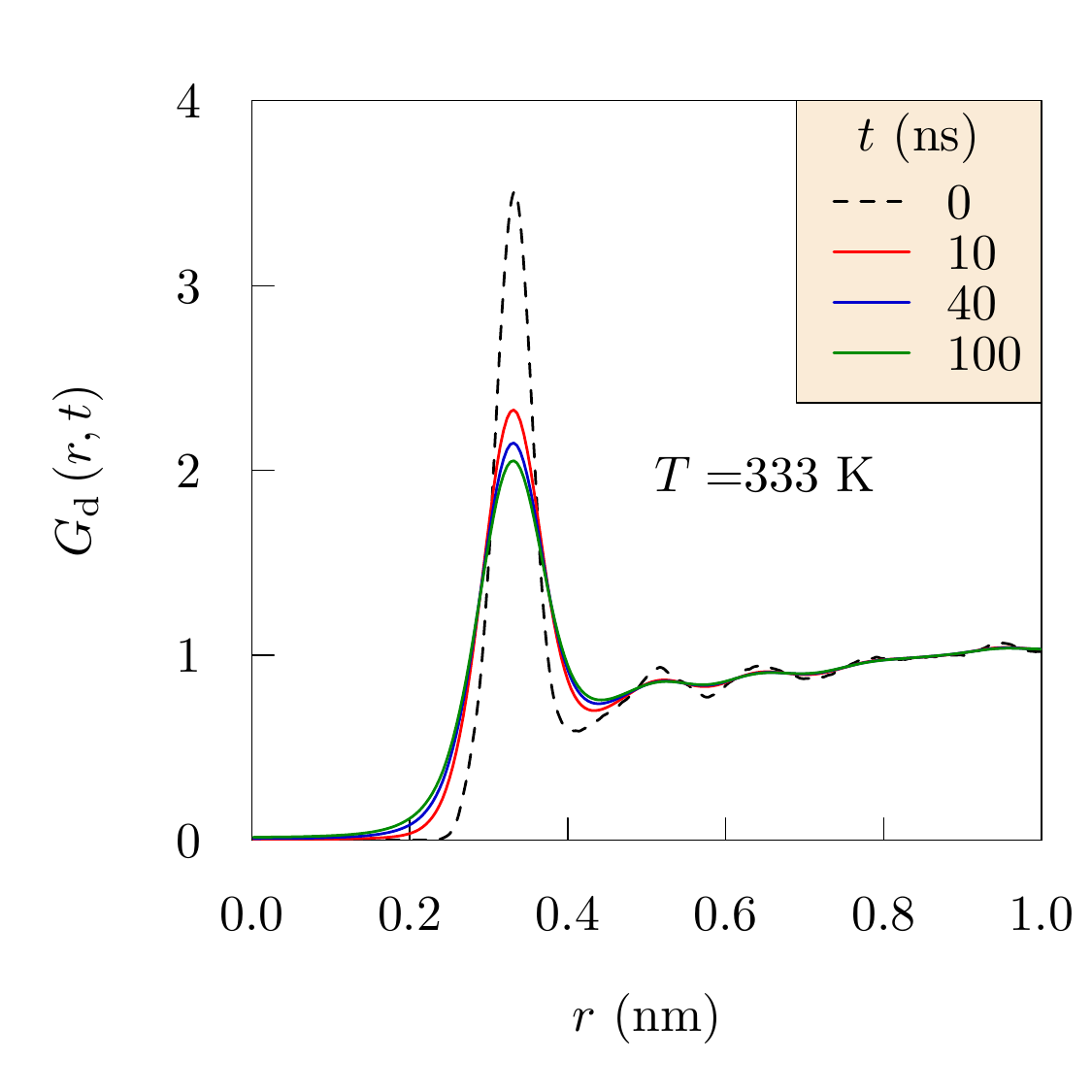}
\includegraphics[width=3.5in]{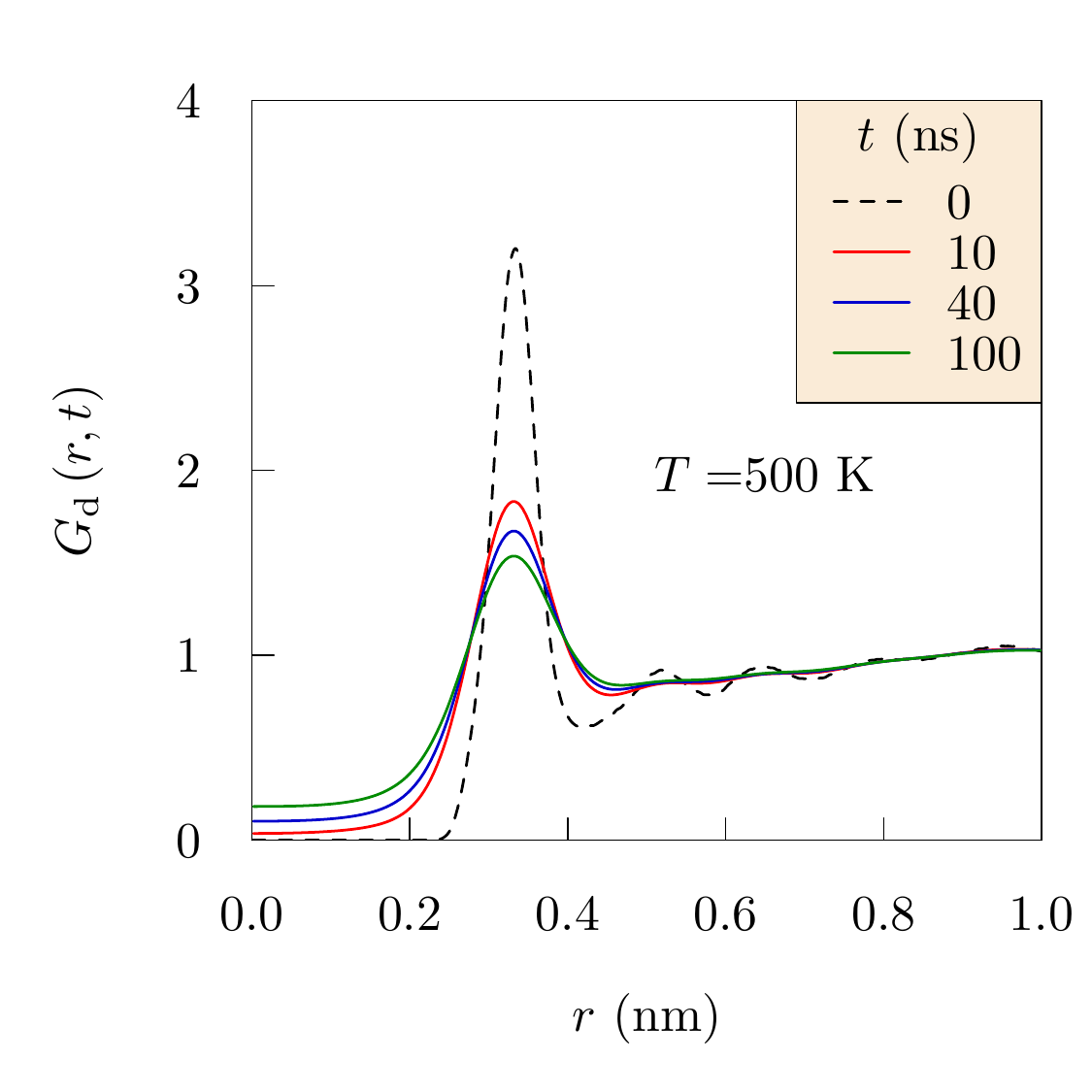}
\caption{The Li-Li radial distribution function  in EDC at
$t=$ 0 (dashed) and the corresponding \emph{distinct} part,
$G_{\text{d}}(r,t)$, of the van Hove function within the Vineyard
approximation. The non-zero population in the core region surrounding  $r
\approx 0$ describes correlation of Li$^+$ jumps, \emph{i.e.,}  refilling a hole left by a Li$^+$ ion with a neighboring Li$^+$ ion. (bottom panel) 
}
\label{vH__333_KSTikZ} \end{figure}

Vineyard's convolution approximation,\cite{Stecki:1970df,YeomansReyna:2000jd}
\begin{equation}
	G_{\mathrm{d}}(r,t)\approx \int \dif^3 r\,^\prime g(r\,^\prime) G_{\mathrm{s}}(|\vec{r}-\vec{r}{}\,^\prime|,t),
\end{equation}
provides an initial characterization for the \emph{distinct} part of the
Li$^+$- Li$^+$ van Hove function (Figure~\ref{vH__333_KSTikZ}). Here,
the convolution of the radial distribution function is made with the
{\em self} part of the van Hove function that is generated by dynamics of
Li$^+$ in EDC. This approximation is consistent with the idea that
$G_{\mathrm{d}}(r,t)$ is a dynamical counterpart to $g(r)$, the radial
distribution function. The non-zero population in the core region
surrounding  $r \approx 0$ describes correlation of Li$^+$ jumps;
that is, refilling a hole left by a Li$^+$ ion with a neighboring
Li$^+$ ion.

\subsection{Vibrational power spectra of Li$^+$}
\begin{figure*}
\includegraphics[width=7in]{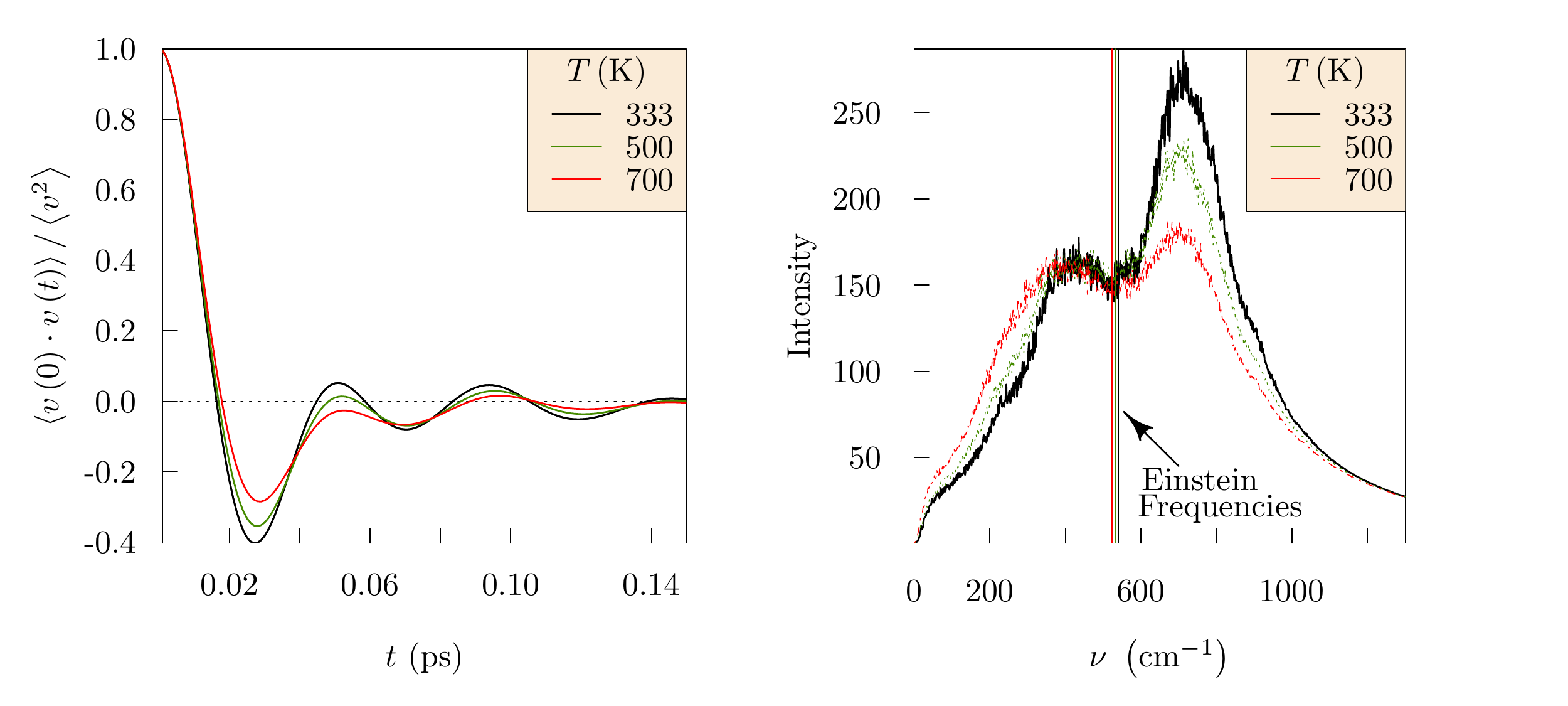}
\caption{Velocity auto-correlation functions (left) for Li$^+$ in EDC at
different temperatures. Power spectra (right) for Li$^+$ in EDC.
Vertical lines near 570 cm$^{-1}$ are Einstein frequencies (Eq.\ref{eq:einstein}) of these
ionic motions. The power spectra identify two prominent vibrational bands, near
400 cm$^{-1}$ and 700 cm$^{-1}$. The intensities of the high frequency
modes diminish with increasing temperature.}
\label{SpectraTikZ} \end{figure*}

\begin{figure*}
\includegraphics[width=3.25in]{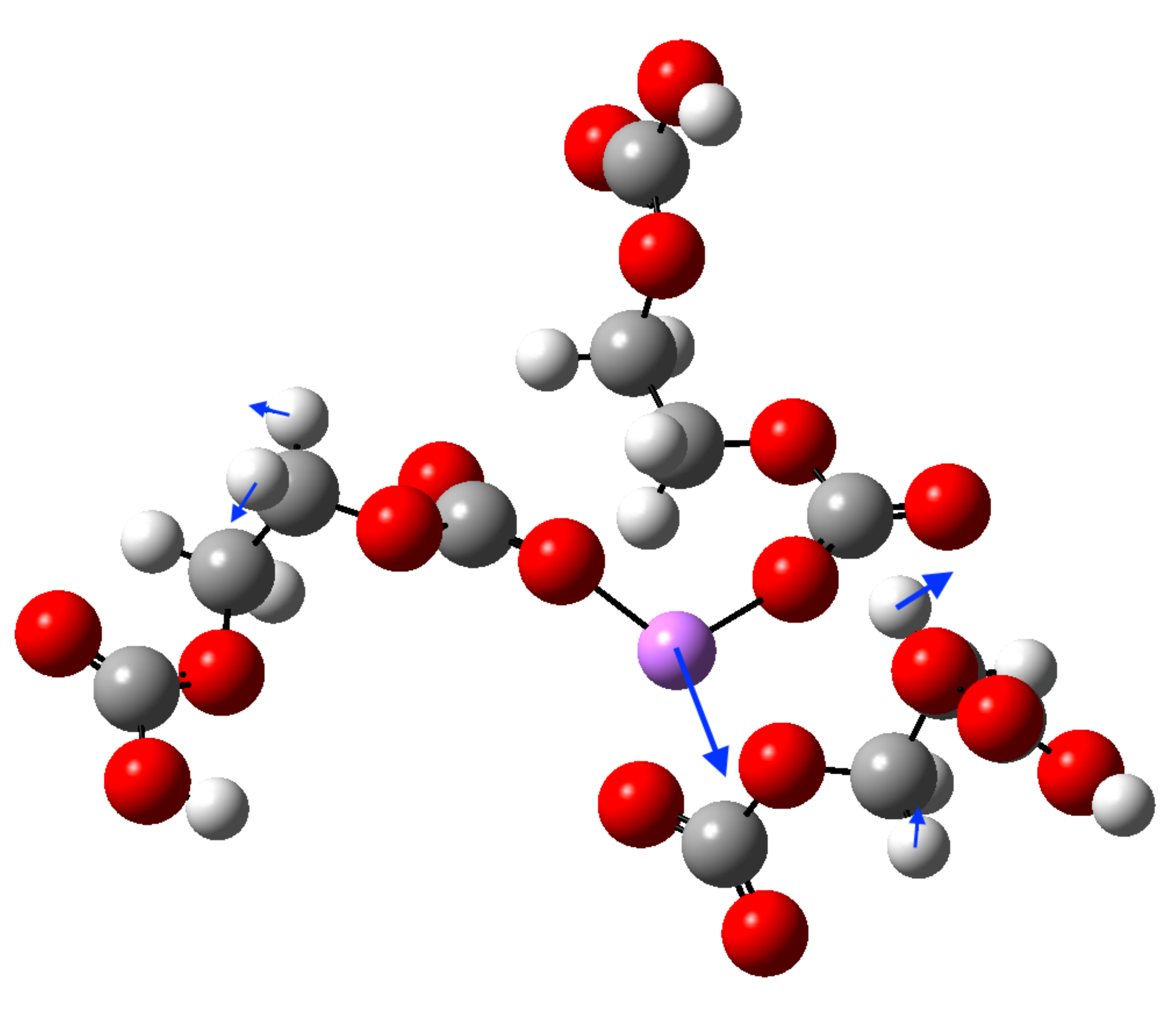}
\includegraphics[width=3.25in]{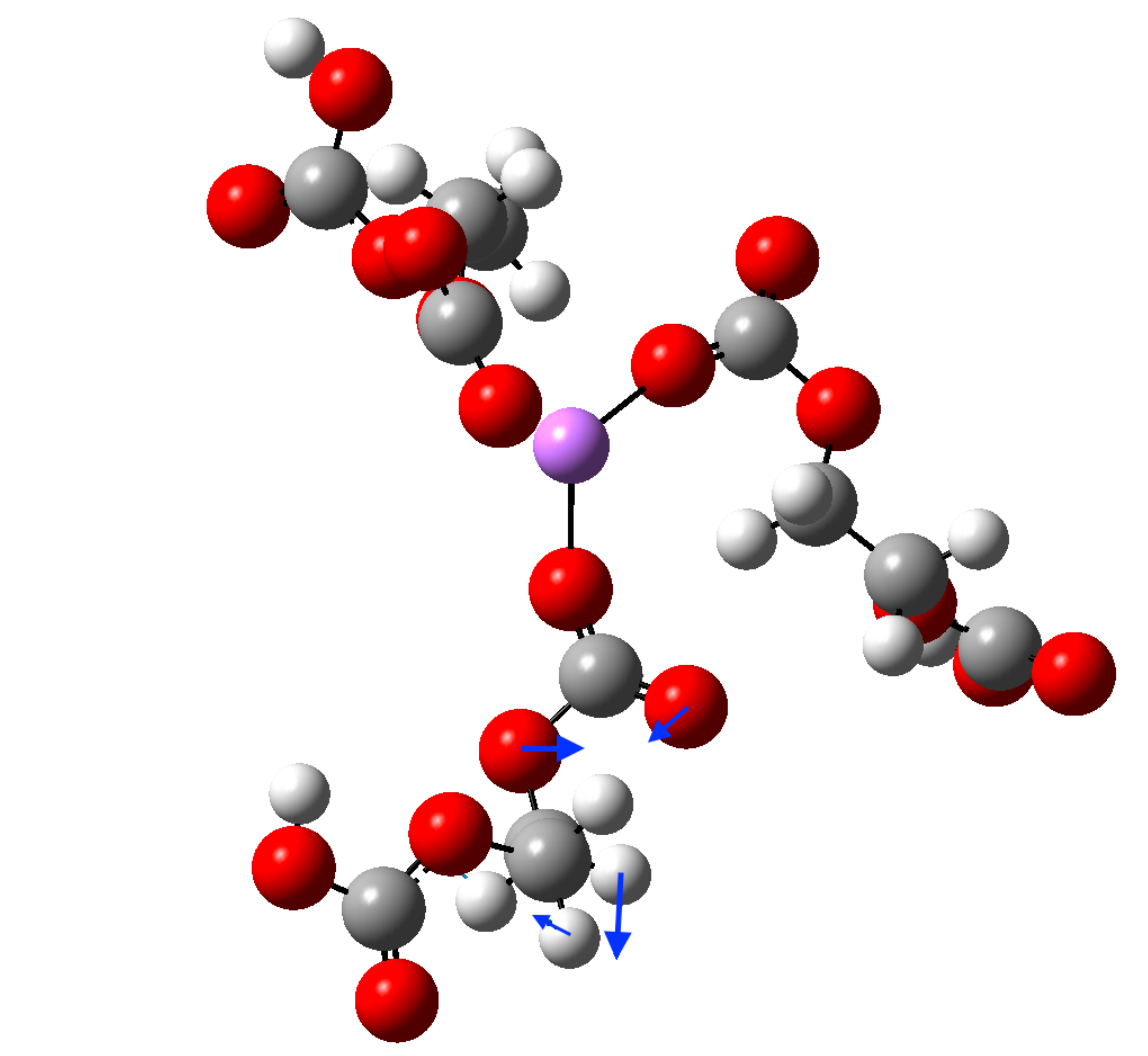}
\caption{Vibrational modes due to prominent Li$^+$ movement (left 436
cm$^{-1}$) and solvent motion around 700 cm$^{-1}$ (right). Li$^+$ is
surrounded by 3 EDC molecules. Adjacent carbonyl and ether oxygen of
the same EDC molecule interact with Li$^+$ (left). The Gaussian09\cite{g09}
software was used for these calculations with the {\tt{b3lyp}}
exchange-correlation density functional and {\tt{6-31+g(d,p)}} basis
set. The structures were sampled from the classical MD simulations and
then the carboxyl groups not coordinating with the Li$^+$ ions were
neutralized by adding hydrogen atoms. The blue arrows indicate the
atomic displacements for these normal mode frequencies.\cite{Rempe:1998}
}\label{fig:gauss_freq}
\label{gaussian_structures} \end{figure*}

The vibrational power spectra are obtained by spectral decomposition of the velocity autocorrelation (Figure~\ref{SpectraTikZ}, left). 
Since we are interested in Li$^+$ transport, this analysis is carried out  for Li$^+$ atoms exclusively. 
This fact distinguishes our results from the FTIR spectrum of EDC molecule that was reported previously.\cite{identify_edc} 
The spectral distribution for Li$^+$ in EDC (Figure~\ref{SpectraTikZ}, right) is bi-modal, 
with a temperature dependence at the higher frequency mode.
Electronic structure calculations, using structures sampled from MD trajectories, confirm these modes and provide a molecular assignment (Figure~\ref{fig:gauss_freq}).
The lower frequency mode (near 400 cm$^{-1}$) corresponds to motion of a
Li$^+$ ion trapped in a cage formed by its nearest neighbors. The higher frequency mode (near 700
cm$^{-1}$) corresponds to Li$^+$ ion picking up the scissoring motion of
a neighboring carbonate group. 

The Einstein frequency ($\nu _{\mathrm{e}}$) is obtained as the coefficient in the quadratic approximation to the velocity autocorrelation function at short times, 
\begin{equation}
  \left\langle \vec{v}\left(t\right)\cdot \vec{v}\left(0\right)\right\rangle/\left\langle v^2\right\rangle \approx 1 - (\nu _{\mathrm{e}}~ 
t)^2
\label{eq:einstein}
\end{equation} In a simple Einstein model, all atoms vibrate with a single frequency. Fittingly, for the bi-modal spectra, $\nu _{\mathrm{e}}$ lies in-between the two significant 
modes.

\section{Conclusions}
Non-polarizable, classical force field parameters were used to study
 transport characteristics  of Li$^+$ in a model SEI layer
composed of EDC.  The structural results in EDC are consistent with
prior studies that use polarizable force fields. An advantage of
non-polarizable force fields is their ready availability in standard
simulation packages and accessibility to  MD studies over microsecond
timescales.  Thus, the dynamical characteristics presented here lay a
basis for careful molecular-scale examination of the mechanism of
transport of Li$^+$ ions in the SEI.

These observations  over microsecond simulation times provide new physical
insights.
Specifically, the results compare the glassy behavior of the
ethylene dicarbonate SEI matrix with the fluid behavior of liquid
ethylene carbonate (Figure~\ref{fig:MSD}). Further, the Li$^+$ MSDs
examined in the nanosecond time intervals distinguish Li$^+$ ion
trapping in cages formed by the EDC matrix. Our results establish the
sizes of the cages and the trapping lifetimes
(Figure~\ref{fig:msdTCompositeTIkZ}), and also the dynamical motions of
the Li$^+$ ions when trapped (Figure~\ref{SpectraTikZ}).  The
vibrational frequency of a trapped ion (about 440~cm$^{-1}$) is
confirmed by electronic structure calculations
(Figure~\ref{fig:gauss_freq}). Our results invalidate a naive Einstein
model of trapped ions that would be plausible otherwise.  The van Hove
correlation functions (Figures~\ref{vH__333_RSTikZ} and
\ref{vH__333_KSTikZ}) provide information for analysis of the
correlation of Li$^+$ jumps.

\section{Methods}
Li$_2$EDC (Figure~\ref{li2edc}) is known to be a dominant component of
the SEI layer in lithium ion batteries involving carbonate solvents.
Although Li$_2$EDC is synthesized in crystalline form, its structure at
the SEI layer is unknown.\cite{Borodin:2013bq,identify_edc} We
constructed a system of 256 Li$_2$EDC moieties for our initial SEI
studies. This system size is identical to previous molecular simulations
performed using polarizable force fields.\cite{Borodin:2013bq,Bedrov:2017ci} For
comparison, we also simulated a single Li$^+$ ion solvated by 249 EC
(Figure~\ref{li2edc}). The GROMACS molecular dynamics simulation
package\cite{Gromacs} was used for all simulations, and the necessary
topology files for EDC and EC were created using non-polarizable
all-atom optimized potentials for liquid simulations (OPLS-AA) force
field parameters.\cite{oplsaa} The partial charges on EC atoms were
adjusted down to 80\% to match experimental transport properties for
EC.\cite{jctc:2016}

The EDC  and Li$^+$ ions were randomly placed into the  simulation cell
and MD simulations were performed at 700, 500 and 333~K. Since EDC ions
are sluggish, configurations from the highest temperature calculation
were used to obtain starting points for further simulations;
cooled down to 500~K and subsequently to 333~K to study moderate
temperature phenomena. Thus, the results presented here are based on
amorphous configurations of the EDC/SEI layer. Although it is unclear
that Li$_2$EDC is crystalline at SEI layer, we have simulated ordered
layers and found that the solvent density and radial distribution
functions are not substantially changed compared with amorphous Li$_2$EDC. The ordered Li$_2$EDC is more
conductive compared to amorphous Li$_2$EDC,\cite{Bedrov:2017ci} but
formation of an ordered SEI structure is unlikely. Therefore, we here
discuss only results of amorphous Li$_2$EDC. 

Periodic boundary conditions  mimic the bulk environment in these
calculations. A Nose-Hoover thermostat\cite{Nose,Hoover} and a
Parrinello-Rahman\cite{barostat} barostat were utilized to achieve
equilibration in the $NpT$ ensemble at 1~atm pressure.  A 200~ns
production run at 700~K was carried out after initial energy
minimization and equilibration steps, then a 250 ns calculation at
500~K. Finally, a 1 $\mu$s trajectory at 333~K temperature was
constructed. Configurations were saved after each 1 ps of those
production runs. A separate 1~ns simulation with a sampling rate of 1~fs
was carried out at each temperature to calculate the time-independent
pair correlation functions discussed below.

\clearpage

\section{Acknowledgement} Sandia National Laboratories is a
multi-mission laboratory managed and operated by National Technology and
Engineering Solutions of Sandia, LLC., a wholly owned subsidiary of
Honeywell International, Inc., for the U.S. Department of Energy's
National Nuclear Security Administration under contract DE-NA-0003525.
This work is supported by the Assistant Secretary for Energy Efficiency
and Renewable Energy, Office of Vehicle Technologies of the U.S.
Department of Energy under Contract No. DE-AC02-05CH11231, Subcontract
No. 7060634 under the Advanced Batteries Materials Research (BMR)
Program. This work was performed, in part, at the Center for Integrated
Nanotechnologies (CINT), an Office of Science User Facility operated for
the U.S. DOE's Office of Science by Los Alamos National Laboratory
(Contract DE-AC52-06NA25296) and SNL.

\section{Author contribution statement}
AM and MIC performed simulations, analysis and prepared all the figures. LRP and SBR wrote the manuscript and all authors reviewed the manuscript.

\section{Additional information}
The author(s) declare no competing financial interests.

\section{Supplementary information}
Diffusion constants, conductivity and a topology file for EDC are reported as  supplementary information.

\end{document}